\documentclass[
aps,prl,
twocolumn,
reprint,
amsmath,amssymb,
superscriptaddress,
notitlepage
]
{revtex4-2}

\usepackage{graphicx}
\usepackage{amsthm}
\usepackage{cancel}
\usepackage{color}
\usepackage{bbold}
\usepackage{wrapfig}
\usepackage{booktabs}
\usepackage{mathtools}
\usepackage{float}
\usepackage{enumitem}
\usepackage{hyperref}
\usepackage{tikz}
\usepackage{comment}
\usepackage{longtable}
\usepackage{multirow}
\usepackage{hyperref}
\usepackage{cleveref}
\usepackage{svg}

\usepackage[normalem]{ulem}

\begin{document}
\global\long\def\pgr{\mathcal{P}_{\text{gr}}}
\global\long\def\pdb{\mathcal{P}_{\text{db}}}
\global\long\def\pov{\mathcal{P}_{\text{ov}}}
\global\long\def\pn{\mathcal{P}_{0}}
\global\long\def\df{d_{\text{f}}}

\title{Synchrony 
for weak coupling in the complexified Kuramoto model}

\author{Moritz Th\"umler}
\affiliation{
  Chair for Network Dynamics,
  Center for Advancing Electronics Dresden (cfaed) and Institute of Theoretical Physics, Technische Universit\"at Dresden, 
  01062 Dresden, Germany
}

\author{Shesha G.M. Srinivas}
\affiliation{Institute of Physics and Material Sciences, Universit\'{e} du Luxembourg}

\author{Malte Schr\"oder}
\affiliation{
  Chair for Network Dynamics,
  Center for Advancing Electronics Dresden (cfaed) and Institute of Theoretical Physics, Technische Universit\"at Dresden, 
  01062 Dresden, Germany
}

\author{Marc Timme}
\affiliation{
  Chair for Network Dynamics,
  Institute of Theoretical Physics, Center for Advancing Electronics Dresden (cfaed) and
  Cluster of Excellence Physics of Life, Technische Universit\"at Dresden, 
  01062 Dresden, Germany
}
\affiliation{
  Lakeside Labs, Lakeside B04b, 9020 Klagenfurt, Austria
}

\begin{abstract}
We present the finite-size Kuramoto model analytically continued from real to complex variables and analyze its collective dynamics.
For strong coupling, synchrony appears through locked states that constitute attractors, as for the real-variable system.
However, synchrony persists in the form of \textit{complex locked states} 
for coupling strengths $K$ below the transition $K^{(\text{pl})}$ to classical \textit{phase locking}. 
Stable complex locked states indicate a locked sub-population of zero mean frequency in the real-variable model and their imaginary parts help identifying which units comprise that sub-population.
We uncover a second transition at $K'<K^{(\text{pl})}$ below which complex locked states become linearly unstable yet still exist for arbitrarily small coupling strengths. 

\end{abstract}

\maketitle

Synchronization, the temporal coordination of two or more state variables, is firmly established as one of the most ubiquitous and most prevalent collective dynamics that emerges across natural and human made networks of interacting units  \cite{pikovsky2001universal, Kurths2007synchronizationOscillatoryNetworks, Schoell2016controlNonlinearSystems, witthaut2017classical}. 
In its simplest realization, the variables of all units become identical over time. 
Other forms of synchrony such as frequency synchronization and phase-locking (see below) may equally prominently emerge \cite{Kurths2007synchronizationOscillatoryNetworks}. 
Emergent synchronization is often essential for system function. For instance, the synchronized dynamics of heart muscle cells enables effective blood circulation \cite{Nitsan2016} and thus life; the phase-locked dynamics among alternating current (AC) phases at different nodes in power grids are necessary for the reliable transfer of electric energy on different scales \cite{Rohden2012synchronizationPowerGrids, Filatrella2008powergridKuramoto-like, witthaut2022collective}. 

\begin{figure}[ht!]
    \centering
    \includegraphics{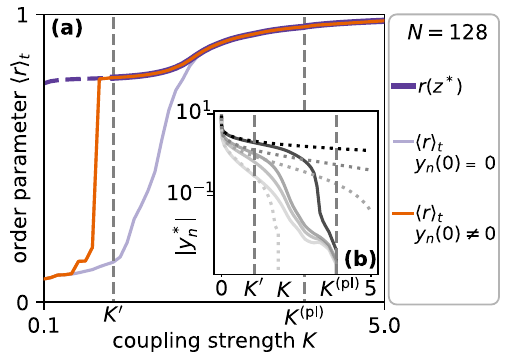}
    \caption{\textbf{Synchrony in the weak coupling regime.}  
    (a) Average order parameters $\langle r \rangle_t$ of a system of $N=128$ complexified Kuramoto units as a function of $K$. The order parameter drops off for $K<K^{(\mathrm{pl})}$ for initial conditions $\boldsymbol{z}(0) = \boldsymbol{x}(0) \in \mathbb{R}^N $ and thus dynamics restricted to the real phase space (light purple). Still, complex locked states $\boldsymbol{z}^*\in \mathbb{C}^N$ persist for complex initial conditions (dark purple), stable for all $K \in (K' , K^{(\mathrm{pl})})$ and unstable for $K<K'$. The dynamics initialized with imaginary parts randomly drawn i.i.d. from a Gaussian distribution, $\mathrm{Im}(\boldsymbol{z}(0)) \stackrel{d}{=} \mathcal{G}(0,10^{-5})$ yields the order parameter $\langle r \rangle_t$ to stay large above $K'$ (orange).
    The inset (b) shows the imaginary parts $y_n^*=\mathrm{Im}(z_n^*)$ at the fixed point against the coupling strength $K$, for four selected units (solid lines), all becoming zero simultaneously at $K^\text{(pl)}$. Asymptotic theoretical approximations (dotted), derived below, smoothly approach these curves. For $K\geq K^\text{(pl)}$, all $y_n^*$ are identically zero. 
    Frequency parameters are i.i.d. random variables from a normalized Gaussian, $\omega_n \stackrel{d}{=}\mathcal{G}(0,1) $.
    }
    \label{fig:large_system}
\end{figure}

The Kuramoto model constitutes a paradigmatic model for collective synchronization phenomena, describing a broad class of weakly coupled strongly attracting limit cycle oscillators \cite{Kuramoto1974nonLinearOscillator, Strogatz2000kuramotoCrawford, acebron2005kuramoto, Rodrigues2016kuramoto, Kuramoto2019reviewPhaseReduction,Lohe_2009}. Its dynamics is given by 
%
$    \frac{\mathrm{d} x_n}{\mathrm{d} t} = \omega_n + \frac{K}{N} \sum_{m=1}^N \sin(x_m - x_n) $
%
for units $n\in\{1,\ldots,N\}$ with phase-like state variables $x_n \in \mathbb{R}$, 
intrinsic frequencies $\omega_n\in\mathbb{R}$ and coupling strength $K\in \mathbb{R}$.
The degree of synchronization among the units is quantified by the mean field order parameter $r(t)\in[0,1]$,  where $r(t)=1$ marks identical synchronization of all units, $x_1(t) = x_2(t) = \ldots = x_N(t)$ in the limit $K \rightarrow \infty$.
The order parameter $r(t)$ is defined via 
\begin{align}
    r(t) e^{\imath \Psi(t)} = \frac{1}{N} \sum_{n=1}^N e^{\imath x_n(t)} \,, \label{eq:mean_order}
\end{align}
with imaginary unit $\imath$. For large coupling strengths, $K>K^\mathrm{(pl)}$, synchronization emerges in terms of phase-locked states with $x_n(t) - x_m(t) = \text{const.}$ for all pairs of units. Decreasing $K$ reduces the alignment of the phases, decreasing the order parameter up to a point $K=K^\mathrm{(pl)}$ below which the phase locked state no longer exists. If the coupling strength decreases further, long-time averages of the order parameter $ \langle r \rangle_t = \lim_{T\rightarrow \infty} \frac{1}{T} \int_0^T r(t) \, \mathrm{d}t$ gradually decrease to $\mathcal{O}(N^{-1/2})$ as $K\rightarrow 0$, see also Fig.~\ref{fig:large_system}a.  
Yet, despite almost half a century of research on this simple-looking model \cite{Strogatz2000kuramotoCrawford,acebron2005kuramoto, Rodrigues2016kuramoto},  several fundamental questions remain open. For instance, essential aspects about the stability of the incoherent and the partially locked states are still unresolved, in particular for finite-size systems \cite{Strogatz2000kuramotoCrawford,acebron2005kuramoto,Mirollo1990, Mirollo2007spectrumPartiallyLockedState, Dietert2017}. Even asking well-posed questions remains a challenge for finite-$N$ systems \cite{Peter2018}.
In this Letter, we take a novel perspective and analytically continue the Kuramoto model to complex variables. We uncover complex locked states that make finite-size ordering phenomena analytically accessible, with implications also for the original, real-variable system.

Analytic continuation of real-valued problems to complex variables has repeatedly shown great success in more profoundly understanding various nonlinear problems. For instance, complexifying the real iterated map $f(x) =x^2+c$ yielded the famous Mandelbrot fractal as the set of complex $c\in \mathbb{C}$ for which the iterated map $z(t+1)=f(z(t))$ does not converge to any invariant set. Moreover, such complexification has catalyzed the identification of radii of convergence for Taylor series of real functions \cite{priestley2003introduction},  conceptually advanced the theory of phase transitions in statistical physics \cite{Lee1952,Yang1952,peng2015experimental}
and initiated the research field of $\mathcal{PT}-$symmetric quantum mechanics \cite{bender1999PTsymmetricQM, Bender2005PTSymmetry}. 

\begin{figure}[ht!]
    \centering
    \includegraphics{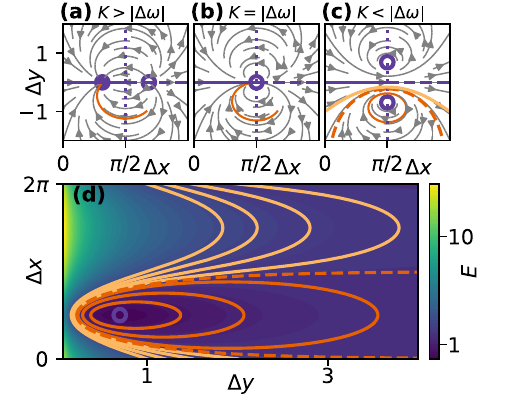}
    \caption{\textbf{Complex locked states for $N=2$.}  
    (a-c) State space vector fields (gray arrows) for (a) strong coupling $K>K^{(\mathrm{pl})}=\Delta \omega$, (b) $K=K^{(\mathrm{pl})}$, and (c) weak coupling in the $N=2$  complexified Kuramoto model. Purple solid disks mark stable, purple open circles unstable or neutrally stable fixed points; purple lines are guides to the eye indicating how fixed points move as $K$ varies (solid: stable, dotted: neutrally stable); red and orange lines show individual sample trajectories, with the dashed line in (c) indicating the separatrix between bound and unbound states. 
    States, with  real parts $\Delta x$ of state differences locked, exist for all $K\in (0, \infty)$. 
    For weak coupling (c), oscillations emerge around locked states at a constant level of $E$, Eq.~\eqref{eq:conservedquantity}. 
    (d) Conserved quantity $E$ determining bounded low energy (red) and unbounded high energy (orange) trajectories. A separatrix (dashed red line) separates low and high-energy regimes at energy $  E_\text{trans.}  =  |\Delta \omega|/K$.
    Further parameters:  $\Delta\omega = 1.0$; in (a): $K = 1.2 \Delta \omega$; in (c) and (d): $K = 0.8 \Delta \omega$.}
    \label{fig:ordertransition}
\end{figure}

Following this line of insights and innovation, we obtain
\begin{align}
    \frac{\mathrm{d} z_n}{\mathrm{d} t} = \omega_n + \frac{K}{N} \sum_{m=1}^{N} \sin(z_m - z_n) \,, \label{eq:model}
\end{align}
for the Kuramoto model, where $z_n = x_n + \imath y_n \in \mathbb{C}$. We observe the dynamical variables $z_n$ in a co-moving reference frame, such that $\langle \omega_n \rangle_n \coloneqq N^{-1}\sum_{n=1}^N\omega_n=0$. Due to the nature of analytic continuation, the state space of the original, real-valued Kuramoto model constitutes an invariant manifold (at $y_n(t)\equiv 0$) embedded into the full complex state space of \eqref{eq:model}.  
We keep the definition \eqref{eq:mean_order} of the order parameter $r(t)$ to study relations of locking phenomena in the complex vs. the original system.

In contrast to the real-valued Kuramoto model, we find that locked states continue to exist for weak coupling, with macroscopic order parameter  $r=\mathcal{O}(1)$, 
indicating a high degree of synchrony among the original real variables $x_n$ (Fig.~\ref{fig:large_system}a). 
Interestingly, below the coupling strength $K^{(\mathrm{pl})}$ above which real phase locked states exist in the original model, the 
complex fixed points $\boldsymbol{z}^*=(z_1^*,\ldots,z_N^*)^\mathsf{T}$ of the system Eq.~\eqref{eq:model} persist and exhibit non-zero imaginary parts $y_n^* \neq 0$ (Fig.~\ref{fig:large_system}b). All $y_n^*$ simultaneously become zero at $K=K^{(\text{pl})}$ and stay zero for stronger coupling. The $\boldsymbol{z}^*$ represent \textit{complex locked states} with all state differences $z_n(t) - z_m(t)$ 
complex constants independent of time.
Our numerical simulations indicate that the fixed points remain attractive in an interval of coupling strengths $K$ below $K^{(\mathrm{pl})}$. As $t\rightarrow \infty$, they are approached 
from initial conditions with small non-zero imaginary parts. 
Crossing a second critical coupling strength $K'<K^{(\mathrm{pl})}$, these locked states continue to exist, yet lose their linear stability (Fig.~\ref{fig:large_system}). Intriguingly, as we explain below also by analytical arguments, these locked states persist for weak coupling and exhibit an order parameter $\langle r(t)\rangle_t \geq 1/\sqrt{2}=\mathcal{O}(N^0)$ also for  $K<K^{(\text{pl})}$ for large $N$ and thus cause strong synchrony of the real parts $x_n$ of the system's state variables, the relevant variables of the original model. This persistence of the locked state stands in stark contrast to the small $\langle r(t)\rangle_t =\mathcal{O}(N^{-1/2})$, non-existence of locked states, and apparent loss of synchrony in the purely real-valued model as $N\rightarrow\infty$.

How does complex synchrony emerge for weak coupling? To address this question, let us first analyze the simplest non-trivial system of $N=2$ complexified Kuramoto units, which effectively reduces Eq.~\eqref{eq:model} to the one-dimensional complex differential equation of the state difference
\begin{align}
    \frac{\mathrm{d} \Delta z}{\mathrm{d} t} = \Delta \omega - K \sin(\Delta z) \,, \label{eq:reduced}
\end{align}
and thus two coupled real differential equations
\begin{subequations}
\begin{align}
    \frac{\mathrm{d} \Delta x}{\mathrm{d} t} &= \Delta \omega - K \sin(\Delta x) \cosh(\Delta y) \label{eq:mod_real}  \\
    \frac{\mathrm{d} \Delta y}{\mathrm{d} t} &=  - K \cos(\Delta x) \sinh(\Delta y) \label{eq:mod_imag} \,,
\end{align}
\end{subequations}
where $\Delta z = z_2 - z_1=\Delta x + \imath \Delta y$  and $\Delta \omega = \omega_2 - \omega_1$. 

For the original real-valued model, i.e., $\Delta y(t)\equiv 0$ for all $t$, phase-locked states with fixed points $\Delta x^*$ exist only for large coupling strengths $K > K^{(\mathrm{pl})} = |\Delta \omega|$, disappearing via a saddle-node bifurcation once $K$ decreases below $K^{(\mathrm{pl})}$. 
The complexified model enables fixed points $\Delta z^*$ for all $K>0$ (Fig.~\ref{fig:ordertransition}).
For $K> |\Delta \omega|$, one stable and globally attractive fixed point coexists with one unstable fixed point on the real axis, as for the real-valued model. These two fixed points bifurcate to a pair of neutrally stable fixed points $\Delta z^*$
with imaginary parts $\Delta y^*\neq 0$ once $K$ drops below $K^{(\mathrm{pl})}=|\Delta \omega|$. Indeed, the complex sine function in Eq.~\eqref{eq:reduced} is an uneven entire function and thus assumes all values in $\mathbb{C}$, such that the fixed point equation $\frac{\partial \Delta z}{\partial t} = 0$ [Eq.~\eqref{eq:reduced}] has solutions for all $K\neq 0$. In  particular, complex locked states exist even for arbitrarily small $K$. An exact analysis detailed in the Supplemental Material \cite{Supplements2022} reveals an energy function 
\begin{align}
    E(\Delta x, \Delta y) = \frac{\Delta \omega}{K} \frac{\cosh(\Delta y)}{ \sinh(\Delta y)} - \frac{\sin(\Delta x)}{\sinh(\Delta y)} \,, \label{eq:conservedquantity}
\end{align}
that is conserved, $\mathrm{d}E/\mathrm{d}t=0$, along trajectories for $K<K^{(\mathrm{pl})}$, establishing that both of the two fixed points at $K<K^{(\mathrm{pl})}$ are indeed neutrally stable nodes (Fig.~\ref{fig:ordertransition}c). Moreover, the analysis uncovers two qualitatively different types of trajectories: low-energy trajectories where the energy lies below a sepa\-ratrix value, $E<E_\text{s}=\Delta \omega/K$, exhibit bounded $\Delta x(t)$ and encircle the nodes; along high-energy trajectories, $\Delta x(t)$ increases without bound (Fig.~\ref{fig:ordertransition}c,d). 
%
%
Next, we demonstrate that synchrony, i.e., macroscopic order, exists for large $N$ with $r=\mathcal{O}(N^0)$ also for $K<K^{\text{(pl)}}$. We do not expect the existence of a conserved quantity such as $E$ in Eq.~\eqref{eq:conservedquantity} for larger $N$, also because numerical simulations indicate convergence to defined states (as illustrated in Fig.~\ref{fig:large_system}) such that phase space volume contracts under the flow. We thus transfer the usage of the fact that the sine functions are entire as above, without invoking a conserved quantity. Specifically, we rewrite the $N$ complex fixed point conditions $\mathrm{d}z_n/\mathrm{d}t = 0$ in the full complexified system Eq.~\eqref{eq:model} in terms of $a_n:=\sin(z_n^*)$ and $\sqrt{1-a_n^2}=\cos(z_n^*)$. Numerically, we observe that the population splits into two groups $\mathcal{N} $ and $\mathcal{P}$,  with $\mathcal{N} = \{n\in\mathbb{N}| \omega_n <0\}$ and $\mathcal{P} = \{n\in\mathbb{N}|\omega_n >0\}$. Further we show analytically, by substituting
\begin{eqnarray}
    a_n \sim & \frac{f_n}{\sqrt{K}} & \text{for } n\in\mathcal{P} \text{ and} \nonumber \\
    a_n \sim & \ \imath \frac{f_n}{\sqrt{K}} & \text{ for } n\in\mathcal{N}, 
    \label{eq:a_n_asymptotics}
\end{eqnarray}
with real $f_n$ and thus $|y_n^*|\sim |\ln(K/(2f_n))|/2$ asymptotically as $K\rightarrow 0$ and
analogously approximating $\sqrt{1-a_n^2}$ asymptotically as $K\rightarrow 0$, into the fixed point conditions $\mathrm{d}z_n/\mathrm{d}t = 0$ yields a consistent asymptotic scaling, indicating that Eq.~\eqref{eq:a_n_asymptotics} is asymptotically exact \cite{Timme2020Scaling} to lowest order (see \cite{Supplements2022} for a detailed step-by-step derivation). 

\begin{figure}[t!]
    \centering
    \includegraphics{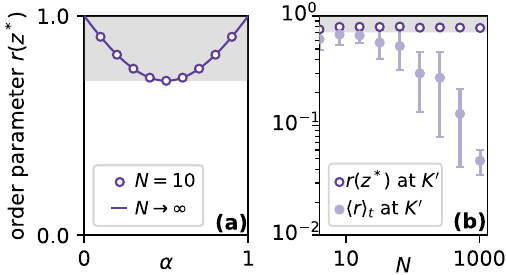}
    \caption{\textbf{Persistent complex synchrony at increasing $N$.}  
    (a) Order parameter $r(\boldsymbol{z}^*)$ vs.~the fraction $\alpha$ of units in population $\mathcal{N}$, asymptotically as $K\rightarrow 0$, as theoretically derived in Eq.~\eqref{eq:orderparameter_asymptotics}. Nine open disks indicate all potential values of $r$ for a small system of $N=10$ units. 
    In the limit $N\rightarrow \infty$, the function $r(\boldsymbol{z}^*(\alpha))$ (solid line) covers a continuous interval $[\frac{1}{\sqrt{2}},1)$ of $r$-values (gray shading).
    (b) Disorder vs. macroscopic order: 
    scaling of $r$ at $K'$ with increasing $N$ for real-variable systems (light purple: $\left<r(t)\right>_t$ evaluated from direct numerical simulations, error bars indicate standard deviation obtained from 20 different realizations of all $\omega_n$, drawn as above) and for the complexified system (dark purple: $r(\boldsymbol{z}^*)$ from the analytical condition $\mathrm{d}\boldsymbol{z}/\mathrm{d}t =0$, error bars smaller than symbol size). Gray shading transferred from (a) indicates that  $r$ stays macroscopic and larger than $1/\sqrt{2}$, also at $K'$.
    }
    \label{fig:macroscopic_order}
\end{figure}

To retrieve the order parameter in the limit of $K\rightarrow 0$, we first disentangle 
\begin{equation}
  a_n = \sin(z_n^*) = \sin(x_n^*)\cosh(y_n^*)+\imath\cos(x_n^*)\sinh(y_n^*)
\end{equation}
separately for  $n\in\mathcal{N}$ and $n\in\mathcal{P}$ employing the asymptotics \eqref{eq:a_n_asymptotics}.  For $n\in\mathcal{P}$, we asymptotically obtain the condition $\text{Im}(a_n)=\cos(x_n^*)\sinh(y_n^*)\rightarrow 0$. Noting that $|y_n^*|>0$, and that indeed $|y_n^*|$ increases with decreasing $K<K^{(\text{pl})}$, the condition becomes $\cos(x_n^*)\rightarrow 0$ such that $x_n^*\rightarrow \pi/2$  and thus $e^{\imath x_n^*}=\imath$ \cite{Footnote}.
Analogously,  for $n\in\mathcal{N}$, we asymptotically require $\text{Re}(a_n)=\sin(x_n^*)\cosh(y_n^*)\rightarrow 0$ and obtain $x_n^*\rightarrow 0$ and thus $e^{\imath x_n^*}=1$ \cite{Footnote}. Now, as the population splits into one sub-population of $\alpha N$ units in $\mathcal{N}$ and one of $(1-\alpha)N$  units in $\mathcal{P}$ as $K\rightarrow 0$, we find
\begin{subequations}
\begin{eqnarray}
    r  = \left|\frac{1}{N}\sum_{n=1}^N e^{\imath x_n^*}\right| 
     & \sim & \left|\alpha + (1-\alpha) \imath \right| \\
     & = & \left(\alpha^2+ (1-\alpha)^2\right)^\frac{1}{2}  = \mathcal{O}(N^0),
\end{eqnarray}
\label{eq:orderparameter_asymptotics}
\end{subequations}
for arbitrarily large $N$, see Fig.~\ref{fig:macroscopic_order}a. As a consequence, the Kuramoto order parameter $r=\mathcal{O}(N^0)$ indicates macroscopic order not only for large $K>K^{(\text{pl})}$ but also asymptotically as $K\rightarrow 0$. It ranges between $r=1/\sqrt{2}$ at $\alpha=1/2$ (population evenly split between $\mathcal{P}$ and $\mathcal{N}$) and $r=1$ at $\alpha=0$ and $\alpha=1$ (almost all units in one sub-population as $N\rightarrow\infty$). 

Logically, the order parameter could scale differently and decrease in the thermodynamic limit as $N\rightarrow\infty$ in some intermediate regime $K\in (K_1,K_2)$ where $K_1>0$ and $K_2<K^{(\text{pl})}$ where the system could cease to show macroscopic order. However, there is no theoretical indication for such a drop and numerical evidence against it for finite systems (see Figs.~\ref{fig:large_system} and  \ref{fig:macroscopic_order}). So by Occam's razor, we conjecture that generically $r=\mathcal{O}(N^0)$ , implying macroscopically ordered collective dynamics emerging as strong synchrony in the form of complex locked states, also for $K\in (K',K^{(\text{pl})})$, see Figure~\ref{fig:macroscopic_order}(b).

\begin{figure}[t!]
    \centering
    \includegraphics{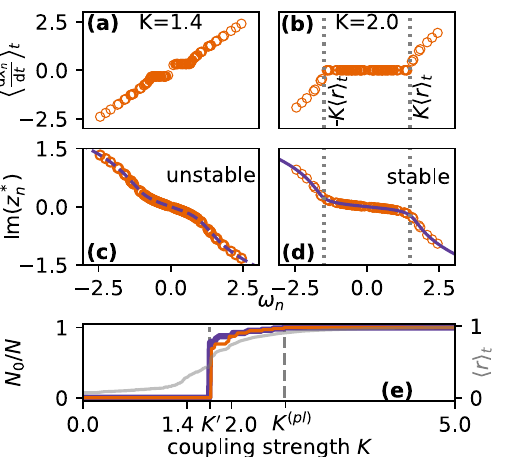}
    \caption{\textbf{Stability of complex locked states indicate real phase-locking.}  \textbf{(a), (b)}, measured mean velocities of $N=128$  real-valued units for two coupling strengths initialized at $y_n(0)=0$. Whereas at $K=2.0>K'$, panel (b), 
    some units lock with zero mean velocity, at $K=1.4<K'$, panel (a), they do not. 
     \textbf{(c),(d)}, imaginary parts of complex locked states obtained from analytical constraint equations \eqref{eq:model}. At $K=2.0$, panel (d), the state is linearly stable (solid line), indicating the existence of locked units in the real-variable model, (b). The points of largest absolute curvature in panel (d) mark the slowest and fastest locked unit, resp.
     At $K=1.4$, the complex locked state is linearly unstable, dashed line, panel (d), coinciding with no zero-velocity locking in the real-variable model, (a).
     \textbf{(e)}, the fraction of units in the locked population $N_0$ varies with coupling strength, once measured from simulations (orange) and once as indicated by the complex locked states for $K>K'$ (purple). The measured averaged order parameter $\langle r \rangle_t$  (gray line) fails to indicate the emergence of real locking.
    Parameters: $\omega_n\stackrel{d}{=}\mathcal{G}(0,1)$, symmetrized (a mild condition widely applied also in the infinite case \cite{Strogatz2000kuramotoCrawford}); see \cite{Supplements2022} for additional details.  
    }
    \label{fig:imaginary_part_story}
\end{figure}

What do complex locked states at $K< K^{(pl)}$ reveal about the dynamics of the original, real-variable Kuramoto model? For general dynamical systems, the stability properties of fixed points considerably influence the local dynamics in their neighborhood. Thus complex locked states $\boldsymbol{z}^*$ close to the real invariant manifold, 
i.e. with small imaginary parts $|y_n^*|$,
impact the dynamics of the real-variable Kuramoto model.
For example, stable complex locked states indicate a locked sub-population at zero mean frequency in the real-variable model and their imaginary parts help quantifying  the size  $N_0 \coloneqq |\{n | \langle \mathrm{d} x_n / \mathrm{d}t \rangle_t = 0\}|$ of that sub-population (Fig.~\ref{fig:imaginary_part_story}).
%

In summary, we have demonstrated that the Kuramoto model analytically continued to complex state space displays strong forms of synchrony with macroscopic order also for $K \rightarrow 0^+$ and as $N\rightarrow\infty$. 

More specifically, the system exhibits three different regimes and thus two transitions as the coupling strength decreases. For strong coupling,   $K>K^{\mathrm{(pl)}}$, real phase-locked states are attractors, as for the original real-variable system.
However, below $K^{\mathrm{(pl)}}$, traditionally marking the transition away from full phase-locking in the real system, synchrony persists, now in the novel form of \textit{complex locked states} with all stationary state variables starting to exhibit non-zero imaginary parts simultaneously. The  Kuramoto order parameter stays macroscopic for large $N$, although it does not account for complex variables $z_n^*$ but only for their real parts $x_n^*$, see Eq.~\eqref{eq:mean_order}. Below a second transition at $K'<K^{\mathrm{(pl)}}$, the complex locked states loose their linear stability, yet continue to persist. In particular, they exist as ordered states with macroscopic $r=\mathcal{O}(N^0)$, even asymptotically as $K\rightarrow 0$.


Prior studies have extended the Kuramoto model to systems with multi-variable units yet not find the unusual ordering phenomena reported above \cite{Chandra2019,OlfatiSaber2006,Zhu2013,Lohe_2009,GU_2007,Jacimovic_2018,Roberts2008linearKuramotoPRE, Muller2021algebraicKuramotoPRE}. A main reason is that these models are all mathematically distinct from the analytic continuation we investigated, because they are topologically distinct, with variables on the $D$-dimensional sphere  \cite{Chandra2019,OlfatiSaber2006,Zhu2013}, algebraically distinct, exhibiting non-Abelian variables (non-commuting matrices) \cite{Lohe_2009,GU_2007,Jacimovic_2018}, or analytically distinct, considering the original phase-like variables $x_n$  as arguments of complex variables $z_n=\exp(\imath x_n)$ of the units  \cite{Roberts2008linearKuramotoPRE, Muller2021algebraicKuramotoPRE}

The complex locked states uncovered above arise from complexification by analytic continuation. They offer a promising alternative perspective for gaining further insights into the order transitions in the real model for finite $N$ and possibly also in the thermodynamic limit $N\rightarrow\infty$ because collective states become traceable through known and exact defining equations for the complex fixed points representing them.
The weak coupling regime $K \in (0,K^{\mathrm{(pl)}})$ of the analytically continued model is of particular interest for large $N$ and demands further explanation, for instance about why and where the second transition at $K'$ occurs as a function of system size $N$ and  realizations of parameters $\omega_n$, compare also \cite{Strogatz2000kuramotoCrawford,Peter2018}. The results reported above specifically suggest that further studying the links between the complex locked states and the locked sub-populations in the original Kuramoto model may be of particular interest, especially as parameters vary.


We have studied the Kuramoto model by analytically continuing the state variables and thus coupling functions, without continuing any of their parameters to become complex. In general, further parameters and models may be complexified to yield a range of new insights for complexified coupled dynamical systems \cite{bender2007complexified, Bender2011tunnelling}, for instance by extending $\omega_n$ or $K$ to be complex and by considering other systems of nonlinearly coupled units such as in the Kuramoto-Sakaguchi model \cite{Omelchenko2012nonuniversalPRL} or (delayed) pulse coupled oscillators \cite{Ashwin_2005}. Such endeavors may 
also conceptually expand our perspective on the nonlinear dynamics of coupled dynamical systems
and intricately wired networks 
\cite{Rodrigues2016kuramoto, Timme2006directedNetworks, dorfler2014synchronization} and thereby initiate research on complexified network dynamics in general.

We thank Philip Marszal for constructive discussions and the German Federal Ministry for Education and Research (BMBF) for support under grant no. 03EK3055F.

\bibliography{refs2.bib}

\begin{thebibliography}{40}
\bibitem [{Foo()}]{Footnote}%
  \BibitemOpen
  \href@noop {} {\bibinfo {title} {{Technically}, we obtain $x_n^*\rightarrow
  \pi/2+ m \pi$ with $m\in \mathbb{Z}$ for {$n\in\mathcal{P}$}; similarly, for
  {$n\in\mathcal{N}$}, we obtain $x_n^*\rightarrow m \pi$ with $m\in
  \mathbb{Z}$. {We} here consider the solutions for $m=0$. {They} contain the
  transition at {$K'$}, and thus exhibit macroscopic order in the regime
  {$K\in[K',\infty)$.}}}\BibitemShut {Stop}%
\end{thebibliography}

\newpage
\end{document}